\documentclass[letterpaper,10pt]{article}
\usepackage{template1}

\usepackage[draft]{hyperref}
\usepackage{amsmath,amssymb,graphicx}
\usepackage{graphicx}
\usepackage{mdwmath}
\usepackage{mdwtab}
\usepackage{comment}

\def\jlt{ J.\ Lightwave\ Technol.\ }
\def\ptl{ IEEE Photon.\ Tech.\ Lett.\ }

\def\opex{ Opt.\ Express }

\def\jlt{ J.\ Lightwave\ Technol.\ }
\def\ptl{ IEEE Photon.\ Tech.\ Lett.\ }

\def\josab{ J.\ Opt.\ Soc.\ Am.\ B }

\def\ol{ Opt.\ Lett.\ }

\begin{document}
\title{Design considerations for multi-core optical fibers in nonlinear switching and mode-locking applications}
\author{Elham Nazemosadat,$^{1,*}$ and Arash Mafi$^{1}$}
\address{$^1$Department of Electrical Engineering and Computer Science, \\ University of Wisconsin-Milwaukee, Milwaukee, WI 53211, USA}
\address{$^*$Corresponding author: nazemosa@uwm.edu}

\begin{abstract}
We explore the practical challenges which should be addressed when designing a multi-core fiber coupler for nonlinear switching or mode-locking applications.
The inevitable geometric imperfections formed in these fiber couplers during the fabrication process affect the performance characteristics of the nonlinear switching device. 
Fabrication uncertainties are tolerable as long as the changes they impose on the propagation constant of the modes are smaller than the linear coupling between the cores. 
It is possible to reduce the effect of the propagation constant variations by bringing the cores closer to each other, hence, increasing the coupling. However, higher coupling translates into a higher switching power which may not be desirable in some practical situations. Therefore, fabrication errors limit the minimum achievable switching power in nonlinear couplers.

\end{abstract}
\maketitle 

\section{Introduction}
Multi-core fiber couplers have recently attracted considerable attention owing to their applications in nonlinear switching, and in particular in all-optical mode-locking of fiber lasers\cite{Winful,Friberg,Proctor,Chao, Multicore,Thomas}.  
A multi-core fiber operates as a linear directional coupler at low optical powers, where power is periodically exchanged amongst neighboring waveguides. Linear coupling is most efficient when the cores are phase matched, or in other words their propagating modes have identical propagation constants~\cite{Saleh}.
At higher optical powers, the nonlinear refractive index of the cores are altered due to nonlinear effects. Consequently, the effective propagation constant of the participating modes are detuned, which reduces the power exchange efficiency between the neighboring cores and as a result the majority of light remains in the launch core~\cite{Jensen}. 
Thereby, the different behaviors a multi-core fiber shows based on the intensity of the input optical power, makes it a suitable device for nonlinear switching and mode-locking applications.

Uncertainties such as fluctuations in the core radius are most likely to happen while drawing a fiber. 
These fabrication errors affect the nonlinear switching  performance of multi-core fiber couplers, and result in a different output compared to what is expected from an ideal coupler with no such uncertainties. 

In this paper, to get a realistic understanding of the performance of multi-core fibers in the presence of fabrication uncertainties, 
the limitations which these errors impose on the performance of a nonlinear coupler are investigated at three levels; 
(1) effects on the propagation constant of the propagating modes $\beta$, 
relative to the core-to-core coupling, 
(2) effects on the first derivative of $\beta$ otherwise known as the group velocity and 
(3) effects on the second derivative of $\beta$ known as the group velocity dispersion (GVD). 
These issues will be addressed in the following sections of the paper.

\section{Uncertainties in fiber fabrication}

In the linear regime, light propagation in each core of a two-core fiber coupler is generally described using the coupled mode theory, 
and can be written as
%%%%%%%%%%%
\begin{align}
\label{eq:ElecField-coupled1}
i~\partial_z 
\begin{pmatrix}
  A_a \\
  A_{b} \\
 \end{pmatrix}
=
\begin{pmatrix}
  \beta_0 & C_{ba} \\
  C_{ab} & \beta_{0} \\
 \end{pmatrix}
\begin{pmatrix}
  A_a \\
  A_{b} \\
 \end{pmatrix}
=\textbf{B A}
\end{align}
%%%%%%%%%%% 
where $A_{a}$ and $A_{b}$ are the amplitudes of the modes in each individual waveguide. It is assumed that 
the propagating modes in both waveguides have identical propagation constants $\beta_0$, and coupling coefficients, $C_{ab}=C_{ba}$. 
Instead of considering the modes of individual waveguides and their couplings, an alternative approach is to study the propagation of the fundamental modes of the complete index profile of the coupler, known as the symmetric and antisymmetric supermodes. The spatial mode profile of the supermodes, $F_{\mu}(x,y)$, can be approximated as a superposition of the guided modes of each individual core as~\cite{Mafi-pcf,MCFvsMMF}
%%%%%%%%%%%
\begin{align}
\label{eq:supermode}
F_{\mu}(x,y)\approx\sum_{\mu=1,2}k_{\mu}\tilde F_{\mu}(x,y),
\end{align}
%%%%%%%%%%%  
where $\tilde F_{\mu}(x,y)$ is the mode profile of the fundamental guided mode propagating in each core and $k_{\mu}$ is the weight factor of the $\mu$th core. 
The eigen-vectors and eigen-values of \textbf{B} provide the weight factors
and propagation constants of the supermodes, respectively. The eigen-vectors of \textbf{B} which are
$ \left( \begin{smallmatrix} 1\\ 1 \end{smallmatrix} \right)$ and $\left( \begin{smallmatrix} -1\\ 1 \end{smallmatrix} \right)$, 
lead to equal excitation of both supermodes
at the fiber input.
However, if due to fabrication uncertainties the propagation constant of one waveguide  slightly changes to $\beta_0+\delta \beta$, the eigenvectors of \textbf{B} change as well, %resulting in  unequal excitation of the supermodes. %In this case, at the coupler input the modes in both waveguides will be excited. 
and consequently the input beam is coupled to the symmetric and antisymmetric supermodes unequally.
This reduces the power exchange efficiency between the two cores and degrades the nonlinear switching performance of the coupler by lowering the modulation depth of the mode-locking device, as also seen in ref.~\cite{MCFvsMMF}. 
In order to avoid this and observe perfect linear switching for low powers, the fluctuations in $\beta_0$ should be much smaller than the coupling between the propagating modes. 
Hence, we can define $\delta \beta \approx C_{ab}$ as the cutoff for the acceptable range of fabrication errors; in this case, around $70\%$ of the power is in one supermode, while the other one only carries $30\%$. 

\begin{figure}[htb]
\centering
\includegraphics[width=2in]{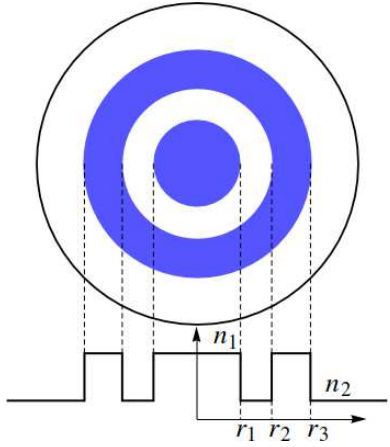}
\caption{Cross sectional view and refractive index profile of the fiber structure.}
\label{fig:fiber}
\end{figure}

To present the aforementioned arguments, we explore the nonlinear switching behavior of a two-concentric-core (TCC) fiber geometry.
This fiber coupler has two concentric cores, consisting of a central
circular core and a surrounding ring core, as shown in Fig.~\ref{fig:fiber}. The refractive index profile of the considered fiber can also be seen in this figure.
Owing to its cylindrical symmetric configuration, this fiber can be fabricated using the same techniques used for conventional fibers, which is more straightforward compared to non-concentric two-core fibers. This simple fabrication process is the main reason for choosing this specific geometry.
One other reason is that the central core of this fiber can be fabricated with the same material and properties of the single mode fibers (SMFs) of the laser cavity, easing the alignment of the coupler with the cavity fibers and reducing the splicing losses~\cite{Cozens}.
%%%%%%%%%%%                                                                                                                                                                             
%%%%%%%%%%%

The fiber profile is designed such that it supports only two zero-angular momentum guided modes referred to as the first and second order guided supermodes.
The electric field in the TCC can be expressed as
the summation of these guided modes as  
%%%%%%%%%%%
\begin{align}
\label{eq:ElectricalFieldEO}
E(x,y,z,t)=\sum_{\mu=1,2}A_{\mu}(z,t)F_{\mu}(x,y)e^{i(\omega_{0}t-\beta_{0,\mu}z)},
\end{align}
%%%%%%%%%%%  
where $A_{\mu}(z,t)$ is the slowly varying envelope of the electric field of the $\mu$th mode with the normalized spatial 
distribution $F_{\mu}(x,y)$ and propagation constant $\beta_{0,\mu}$.
The generalized nonlinear Schr\"{o}dinger equation (GNLSE) describing the longitudinal evolution of 
a pulse in this fiber coupler can be written as~\cite{Poletti,Mafi}
%%%%%%%%%%%
\begin{align}
\label{eq:master}
\nonumber
\dfrac{\partial A_{\mu}}{\partial z}&=
i~\delta \beta^{(0)}_{\mu}A_{\mu}-\delta \beta^{(1)}_{\mu}\dfrac{\partial A_{\mu}}{\partial t}
-i\dfrac{\beta^{(2)}_{\mu}}{2}\dfrac{\partial^2 A_{\mu}}{\partial t^2}\\
&+i(\dfrac{n_2\omega_0}{c})\sum_{\nu,\kappa,\rho=1,2}f_{\mu\nu\kappa\rho}A_\nu A_\kappa A^\ast_\rho,
\qquad \mu=1,2,
\end{align}
%%%%%%%%%%%
The indices can take the value of $1$ or $2$ corresponding to first and second order supermodes. $\beta^{(1)}_{\mu}$ denotes the group velocity, and
$\beta^{(2)}_{\mu}$ is the GVD parameter of the $\mu$th mode.
We define
$\delta\beta^{(0)}_{1}=-\delta\beta^{(0)}_{2}=\beta^{(0)}_{1}-\beta^{(0)}_{ref}$ and 
$\delta\beta^{(1)}_{1}=-\delta\beta^{(1)}_{2}=\beta^{(1)}_{1}-\beta^{(1)}_{ref}$, where 
$\beta^{(i)}_{ref}=(\beta^{(i)}_{1}+\beta^{(i)}_{2})/2$ for $i=~0,1$. The nonlinear 
index coefficient of the fiber is defined by $n_2$ and $\omega_0$ is the carrier frequency.
The nonlinear coupling coefficients $f_{\mu\nu\kappa\rho}$ are given by
%%%%%%%%%%%
\begin{align}
\label{eq:nonlinear_coefficient}
f_{\mu\nu\kappa\rho}=\iint{dx dy}~F^\ast_\mu F_\nu F_\kappa F^\ast_\rho,
\end{align}
%%%%%%%%%%%
where the mutually orthogonal spatial profiles are assumed to be normalized according to
$\iint F^2_1~dxdy=\iint F^2_2~dxdy=1$. 

\begin{figure}[htb]
\centering
\includegraphics[width=2.5in]{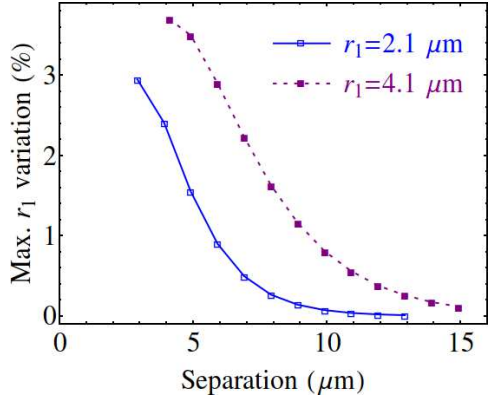}
\caption{The maximum acceptable fabrication errors in the central core radius as a function of the separation between the cores.}
\label{fig:variation}
\end{figure}

In the two different TCC fiber couplers studied here, in order to have maximum linear coupling, the inner and outer radii of the ring core are designed such that the propagation constant of its fundamental mode matches that of the central core. 
The radius of the central core is altered from its original value to the cutoff value, where one supermode carries 70\% of the input power and defined that value as the maximum acceptable fabrication errors. 
The core size of the central core is the only geometrical parameter altered in this paper, because it has been shown that the performance of this fiber is more sensitive to the variation of the central core radius compared to that of the ring core~\cite{Nunes}.
The radius of the central core and the refractive index of both cores in coupler A are $r_1=2.1~{\rm \mu m}$ and $n_{1}=1.465$, respectively, whereas the parameters of coupler B are $r_1=4.1~{\rm \mu m}$ and $n_{1}=1.451$. The operating wavelength is $\lambda_0=1550$~nm, and the cladding refractive index for both couplers is assumed $1.444$ (corresponding to pure silica). 
Fig.~\ref{fig:variation} shows how the tolerance to fabrication errors in these two couplers varies as a function of the separation between the two cores, where the separation is defined as the gap between the two cores ($r_2 - r_1$ in Fig.~\ref{fig:fiber}). 
As the separation between the ring core and the central core increases, the overlap of the fields reduces, which leads to a smaller coupling and a larger beat length.
According to the results, the tolerance to fabrication errors reduces as the ring core is placed further away from the center core.  Hence it is preferred to design the coupler such that the separation between the two cores is small, at the price of a higher switching power.
Typical fabrication errors in the core radius are around $\pm1\%$, thus according to Fig.~\ref{fig:variation}, the maximum allowable separation between the two cores in coupler A is around $6~\mu m$, 
while in coupler B separations up to $10~\mu m$ are acceptable.
%%%%%%%%%%%                                                                                                                                                                      
%%%%%%%%%%%
\section{Nonlinear switching behavior}
\begin{figure}[htb]
\centering
\includegraphics[width=2.5in]{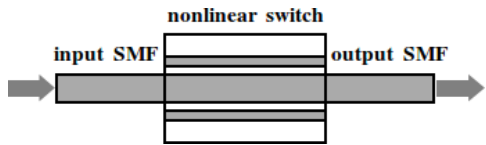}
\caption{An all-fiber nonlinear switching device composed of a TCC coupler placed in between two SMFs.}
\label{fig:coupler}
\end{figure}

The TCC fiber can be used as an all-fiber mode-locking device, referred to a saturable absorber, as shown in Fig.~\ref{fig:coupler}, where the beam is injected into the central core of the TCC coupler using an input
SMF and is collected at the output from the same core, using another SMF. 
To minimize the splicing loss between the cavity and the TCC,
the central core is considered identical to the cavity SMFs. 
Given that the injected beam from the input 
SMF is in the form of an azimuthal symmetric Gaussian beam, by aligning the central core of the TCC with the input SMF core, only the zero angular momentum modes of the TCC fiber can be excited~\cite{MafiMMI1}, which is the case considered in this paper.
%%%%%%%%%%%                                                                                                                                                                             
%%%%%%%%%%%

When the noise of the gain medium in the laser cavity is transmitted through the TCC fiber coupler with a length equal to the coupler's half-beat-length $L_{hb}$, low intensity noise is linearly coupled to the ring core and experiences a high loss, while high intensity noise remains in the launch core, resulting in a high intensity contrast. This leads to the formation of optical pulses. The power-dependent transmission feature of the coupler can be used for mode-locking fiber lasers~\cite{Winful,Proctor}.  
In modern lasers, the saturable absorber is a critical device only for allowing the pulse to build up from noise and start the lasing process, while the pulse shaping is achieved through using a narrow spectral filter~\cite{Chong}. 
Therefore, it would be reasonable to study the performance of the device for nanosecond pulses, which are pertinent to the starting process. 
 
A hyperbolic secant optical pulse with
a pulse width of $t_0$, is launched into the nonlinear coupler through the input SMF.
The nearly Gaussian spatial mode profile of the input beam excites the supermodes of the coupler. The excitation power related to each supermode can be found using the overlap integral of that particular supermode with the input beam. The pulse propagation inside the nonlinear coupler was analyzed by solving Eq.~\ref{eq:master} using the split step Fourier method.
The collected output power from the coupler is $P_{out}=|A_{c}|^2$, where $A_{c}$ is the field amplitude of the center core, obtained by
%%%%%%%%%%%
\begin{align}
\nonumber
A_c(z,t)&=A_1(z,t)\iint F^\ast_cF_1 dxdy\\
&+A_2(z,t)\iint F^\ast_cF_2 dxdy,
\label{eq:core-field-amplitude}
\end{align}
%%%%%%%%%%%
where $F_{\rm 1}$, $F_{\rm 2}$ and $F_{\rm c}$ are the spatial mode profiles of the two supermodes of the TCC coupler and that of the propagating mode of the central core, respectively. The finite element method was used to calculate these profiles~\cite{Lenahan}. $F_{\rm c}$ is calculated for the coupler in the absence of the outer ring.
The transmittance is defined as the energy at the output of the center core divided by the total energy injected into the coupler, given by
%%%%%%%%%%%
\begin{equation}
\tau=\dfrac{\int_{-\infty}^{+\infty} |A_c(L_{hb},t)|^2 dt}{\int_{-\infty}^{+\infty} (|A_1(0,t)|^2+|A_2
(0,t)|^2)dt},
\label{eq:tauSMFGIMFSMF}
\end{equation}
%%%%%%%%%%%

%%%%%%%%%%                                                                                                                                                                          
\begin{figure}[htb]
\centering
\includegraphics[width=2.5in]{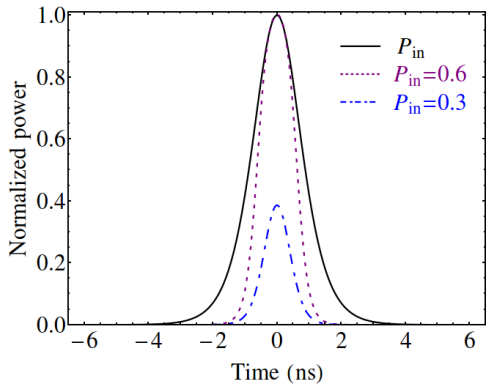}
\caption{The normalized pulse power at the output of the center core for different input peak powers (units are MW) when the pulse width is $t_0=1~\rm {ns}$. 
A similar behavior is observed for picosecond pulses, but we considered nanosecond pulses because they are more relevant for the start of a lasing process.
}
\label{fig:Pout}
\end{figure}
\begin{figure}[htb]
\centering
\includegraphics[width=2.5in]{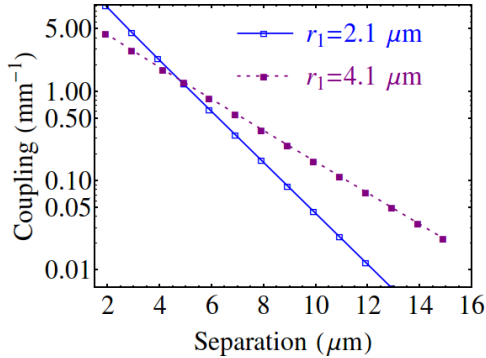}
\caption{The coupling among the two cores as a function of the separation between them in logarithmic scale.}
\label{fig:coupling}
\end{figure}
\begin{figure}[htb]
\centering
\includegraphics[width=2.5in]{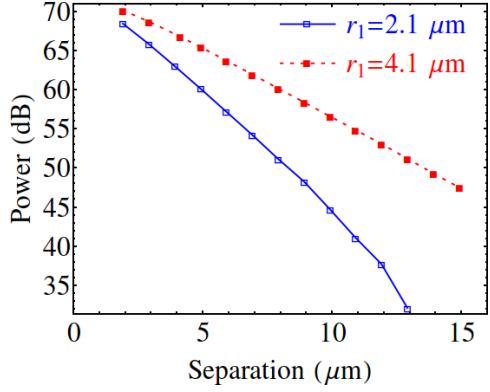}
\caption{The approximate power required for nonlinear switching in couplers A and B as a function of the separation between the cores.}
\label{fig:P_switching}
\end{figure}

%%%%%%%%%%%
In Fig.~\ref{fig:Pout}, the normalized input pulse and the output pulses corresponding to different peak input pulse intensities are displayed for coupler A. 
The fiber is assumed to be made of silica and its parameters, with referring to Fig.~\ref{fig:fiber}, are $r_1=2.1$, $r_2=8$ and $r_3=9.95~{\rm \mu m}$. These parameters are chosen such that the coupler's tolerance to fabrication errors is around $\pm1\%$ as shown in Fig.~\ref{fig:variation}. 
It can be observed that as the input power increases, a larger portion of the pulse energy remains in the central waveguide. 
For the parameters considered here, a switching power in the order of 0.5~MW is required.
Transition from low transmission to high transmission occurs when the nonlinear interaction of each mode with itself is comparable to the linear core-to-core coupling ; therefore, a fair estimate for obtaining the power threshold for nonlinear switching in an ideal coupler without any fabrication errors is $(n_2\omega_0/c)f_{1111} P_0\approx C$, where $C=\delta\beta_{0,1}$ is the linear coupling coefficient among the two propagating modes. Hence, a lower switching power requires a coupler with a large nonlinear coefficient and a small linear coupling coefficient. In Fig.~\ref{fig:coupling} it can be seen how the linear coupling coefficient, $C$, varies as a function of the separation between the cores of couplers A and B. To better observe the difference between the $C$ parameter of these two couplers, the figure is plotted in the logarithmic scale.  As observed in the figure, for small separations, the coupling coefficient in coupler A is larger than coupler B, while the vice versa occurs for larger separations. Considering a fixed separation, the effective area of coupler A is less than that of coupler B, resulting in a larger nonlinear coefficient. For the specific couplers considered here, the nonlinear coefficient of coupler A is approximately 3 times that of coupler B for all separations. The combination of linear coupling and nonlinearity leads to a lower switching power in the coupler with the smaller central core radius, for all separation values, as can be seen in Fig.~\ref{fig:P_switching}. 
To reduce the switching power, one can increase the separation between the two cores and reduce the coupling. However, as mentioned in the previous section, such a coupler with a large separation between the two cores requires high fabrication precision in order to operate properly in the linear regime.
%%%%%%%%%%%                                                                                                                                                                      
%%%%%%%%%%%
%%%%%%%%%%%                                                                                                                                                                      
%%%%%%%%%%%

In Fig.~\ref{fig:P_switching} sensitivity of the coupler transmission has been studied against $\pm1\%$ variations in the radius of the central core. As observed, the modulation depth of the nonlinear switch is reduced in both cases. When there are no variations in the center core, both supermodes are excited equally, while in $\delta_{r_1}=1\%$, most of the power is carried by the first order supermode and vice versa for the $\delta_{r_1}=-1\%$ case. This unequal excitation of the modes is the reason behind the lower quality of the transmission curves of these two cases compared to the error-free case. 
$\delta_{r_1}=1\%$ variations have an advantage of shifting the transmission curve to lower powers whereas, $\delta_{r_1}=-1\%$ variations increase the power threshold of the device which is not desirable. 
Larger amounts of error reduce the modulation depth furthermore, thus there is a compromise between the fabrication error and the modulation depth of the device. 
In cases where a lower modulation depth is sufficient to start the lasing process of the laser, one can deliberately increase the central core radius slightly from the designed value, and hence reduce the required switching power.
%%%%%%%%%%%                                                                                                                                                                      
\begin{figure}[htb]
\centering
\includegraphics[width=2.5in]{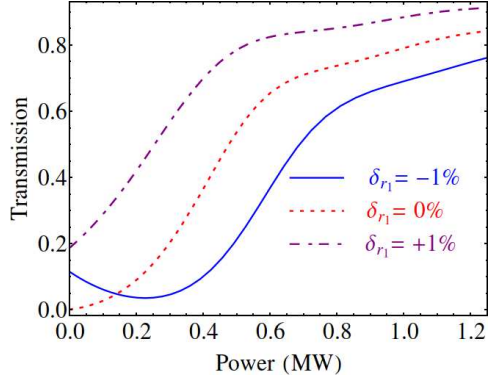}
\caption{The relative power transmission of coupler A is plotted as a function of the input power with variations in the central core radius.}
\label{fig:Transmission}
\end{figure}
%%%%%%%%%%%                                            
\section{Group velocity}
The two supermodes have different group velocities; thus the pulses carried by these two modes walk-off due to traveling at different speeds and after a certain propagation distance known as the walk-off length, given by $L_w=t_0/(\beta_{1,1}-\beta_{1,2})$, the two pulses won't have any overlap and hence will not interact. 
Given that the length of the coupler is chosen based on its half-beat-length, $L_{hb}$, to avoid the walk-off effect, while designing a fiber coupler one should take note that the minimum pulse width which can propagate through the coupler without experiencing walk-off is $t_0=L_{hb}\times (\beta_{1,1}-\beta_{1,2})$.
For the considered coupler length in this paper, the minimum acceptable pulse width which won't experience walk-off is $t_0\approx14.7~$fs.  

\section{Group velocity dispersion}
When the two cores are strongly coupled and the separation between them is not large, the GVD and also dispersion are comparable to that of SMF-28 and fluctuations due to fabrication imperfections do not affect the dispersion significantly. %Thus, as mentioned above it is preferred to use small separations. 
It is interesting to note that while for large separations the dispersion in the specific fiber geometry considered here increases rapidly~\cite{Thyagarajan}, possible fabrication fluctuations reduce the dispersion considerably. Hence, fabrication uncertainties not only are not a limiting factor in terms of the waveguide dispersion, but in fact help reduce the dispersion as well. 
In addition, for the pulse widths considered here, the dispersion length of the fiber is much longer than the coupling length, so even high amounts of dispersion do not affect the coupler's performance considerably.

\section{Conclusions}
From a practical point of view, we have studied the design considerations for multi-core fiber couplers used in nonlinear switching and mode-locking applications.
The effect of fabrication imperfections on the performance of a two core fiber is investigated, which can be generalized to all multi-core couplers.  It is observed that there is a compromise between the fabrication error and the required nonlinear switching power and also the modulation depth of the device.
However, to take advantage of these errors, it is shown that if the central core radius is slightly larger than the optimum value, the switching power can be reduced down to around half at the expense of a lower modulation depth.
It should also be pointed out that
to decrease the power threshold, the coupler can be fabricated from material with a higher nonlinearity compared to silica~\cite{Chalcogenide}.

\section*{Acknowledgement}
The authors acknowledge support from the Air Force Office of Scientific Research under Grant FA9550-12-1-0329.

%%%%%%%%%%%%%%%%%%%%%%%%%%%%%%%%%%%%%%%%%%%%%%%%%%%%%%%%%%%%%%%%%%%%%%%%%%%%%%%%%%%%%%%%%%%%%%%%%%%%%%%%%%%%%%%%%%%%
\end{document}